\documentclass[10pt,english,two column]{IEEEtran}
\renewcommand{\baselinestretch}{1}
\usepackage[T1]{fontenc}
\usepackage[latin9]{inputenc}
\usepackage{float}
\usepackage{amsmath}
\usepackage{amsthm}
\usepackage{amssymb}
\usepackage{graphicx}
\usepackage{multirow}
\usepackage{xcolor}
\usepackage{subcaption}
\usepackage{multicol}
\usepackage{algorithmic}
\usepackage[font=footnotesize]{caption}

\makeatletter

\floatstyle{ruled}
\newfloat{algorithm}{tbp}{loa}
\providecommand{\algorithmname}{Algorithm}
\floatname{algorithm}{\protect\algorithmname}

\theoremstyle{plain}

\theoremstyle{plain}

\ifCLASSINFOpdf
\else
\fi
\usepackage{babel}

\makeatother

\usepackage{babel}
\providecommand{\propositionname}{Proposition}
\providecommand{\theoremname}{Theorem}

\begin{document}

\title{\huge{Learning Robust Beamforming for MISO Downlink Systems}}

\author{Junbeom Kim, \textit{Student Member, IEEE}, Hoon Lee, \textit{Member, IEEE}, and Seok-Hwan Park, \textit{Member, IEEE} {\vspace{-6mm}} \thanks{This work was supported by the National Research Foundation of Korea (NRF) grants funded by the Korea government (MSIT) (No. 2019R1F1A1060648, 2019R1A6A1A09031717, 2021R1C1C1006557).

J. Kim and S.-H. Park are with the Division of Electronic Engineering, Jeonbuk
National University, Jeonju 54896, South Korea (email: \{junbeom, seokhwan\}@jbnu.ac.kr).

H. Lee is with the Department of Smart Robot Convergence and Application Engineering and the Department of Information
and Communications Engineering, Pukyong National University, Busan 48513, South Korea (e-mail: hlee@pknu.ac.kr).}}
\maketitle
\begin{abstract}
This paper investigates a learning solution for robust beamforming optimization in downlink multi-user systems. A base station (BS) identifies efficient multi-antenna transmission strategies only with imperfect channel state information (CSI) and its stochastic features. To this end, we propose a robust training algorithm where a deep neural network (DNN), which only accepts estimates and statistical knowledge of the perfect CSI, is optimized to fit to real-world propagation environment. Consequently, the trained DNN can provide efficient robust beamforming solutions based only on imperfect observations of the actual CSI. Numerical results validate the advantages of the proposed learning approach compared to conventional schemes.
\end{abstract}

\vspace{-2mm}

\theoremstyle{theorem}
\newtheorem{theorem}{Theorem}
\theoremstyle{proposition}
\newtheorem{proposition}{Proposition}
\theoremstyle{lemma}
\newtheorem{lemma}{Lemma}
\theoremstyle{corollary}
\newtheorem{corollary}{Corollary}
\theoremstyle{definition}
\newtheorem{definition}{Definition}
\theoremstyle{remark}
\newtheorem{remark}{Remark}

\vspace{-2mm}
\section{Introduction}
Multi-antenna beamforming techniques have been regarded as key enablers of wireless communication systems thanks to the capability of mitigating inter-user interference \cite{Bjornson-et-al:SPM14}. To find an efficient beamforming solution, a base station (BS) needs to get a perfect access to channel state information (CSI). With the perfect CSI at hand, the BS can determine the beamforming solution based on existing optimization algorithms such as weighted minimum mean squared error (WMMSE) algorithm \cite{Christensen-et-al:TWC08}. However, the practical CSI acquisition steps incur inevitable errors in observable CSI. Due to the model mismatch, beamforming methods developed for an ideal perfect CSI cannot capture the impact of erroneous CSI and degrade the system performance.

Robust beamforming techniques have been investigated for computing the beamforming vectors only with the imperfect CSI knowledge \cite{Wang-et-al:WCL12,Choi-et-al:TWC20}. Existing approaches focus on deriving new performance metrics that can compensate for the mismatch between the observable imperfect CSI and the actual propagation environment. The channel estimation error fundamentally entails intractable formulas in analyzing the system performance, and thus the conventional methods mostly rely on inaccurate approximations and assumptions. For instance, the signals obtained through the CSI error are typically treated as additive noise \cite{Wang-et-al:WCL12} and interference \cite{Choi-et-al:TWC20}, respectively, although they indeed contain the intended data. Also, reference \cite{Choi-et-al:TWC20} maximizes the lower bounds of the weighted sum rate, possibly posing the optimality loss.

The drawbacks of the conventional studies are stemmed from model-based optimization approaches that require analytical formulations for the performance metric. To overcome this challenge, this letter presents a data-driven deep learning (DL) solutions for the robust beamforming optimization. The DL techniques have recently made significant progress for handling non-convex optimization problems in wireless communications \cite{Sun-et-al:TSP18}-\cite{Lee-et-al:JSAC19}. The authors in \cite{Xia-et-al:TC19} laid the cornerstone on the DL-based optimization of downlink transmit beamforming. Efficient construction strategies of deep neural networks (DNNs) have been presented in \cite{Kim-et-al:WCL20}. These studies are, however, confined to the ideal perfect CSI assumption and cannot be viable in practical wireless systems.

This letter proposes a DL approach that determines robust beamforming solutions \cite{Wang-et-al:WCL12, Choi-et-al:TWC20, Shen-et-al:TVT14} that compensate for the effect of the channel estimation error by using the imperfect CSI knowledge. We design a DNN to exploit observable information of the channels, i.e., the estimated CSI and its error statistics, so that it can compute the beam solution only using the nominal CSI. To narrow the gap between performance of the perfect and estimated channels, a robust training strategy is proposed which optimizes the DNN both with the actual CSI and artificially generated estimation error. Unlike existing approaches \cite{Xia-et-al:TC19, Kim-et-al:WCL20} which only use the perfect CSI in the training step, the proposed robust DL method helps the DNN learn the stochastic features of the actual channels via numerous realizations of the estimated CSI. Numerical results verify the robustness of the proposed DL schemes over conventional algorithms.

\vspace{-2mm}
\section{{System Model and Problem Formulation}\label{sec:System-Model}}
This section describes a system model for multi-user MISO broadcast channels in which downlink data transmissions of $K$ single-antenna user equipments (UEs) are simultaneously carried out by a BS with $M$ transmit antennas. The received signal $y_{k}$ for UE $k\in\mathcal{K}\triangleq\{1,\cdots,K\}$ is expressed as $y_{k} = \mathbf{h}_{k}^{H}\mathbf{v}_{k}s_{k} + \mathbf{h}_{k}^{H}\sum_{j\in\mathcal{K}\setminus\{k\}}\mathbf{v}_{j}s_{j} + z_{k}, \label{eq:yk}$
where $\mathbf{h}_{k}\in\mathbb{C}^{M}$ denotes the channel vector from the BS to UE $k$, $\mathbf{v}_{k}\in\mathbb{C}^{M}$ stands for the beamforming vector for UE $k$, $s_{k}\sim \mathcal{CN}(0,1)$ indicates the data signal intended to UE $k$, and $z_{k}\sim\mathcal{CN}(0,1)$ is the additive noise. The sum power constraint at the BS is imposed as $\sum_{k\in\mathcal{K}}\|\mathbf{v}_{k}\|^{2}=P$ with the power budget $P$. Defining $\mathbf{h}\triangleq\{\mathbf{h}_{k}:\forall k\in\mathcal{K}\}$ and $\mathbf{v}\triangleq\{\mathbf{v}_{k}:\forall k\in\mathcal{K}\}$, the achievable rate of UE $k$ is written by
\begin{align}
R_{k}(\mathbf{h}_{k}, \mathbf{v})=\text{log}_{2} \bigg(1 + \frac {|\mathbf{h}_{k}^{H}\mathbf{v}_{k}|^{2}}{\sum\nolimits _{j\in\mathcal{K}\setminus\{k\}}|\mathbf{h}_{k}^{H}\mathbf{v}_{j}|^{2}+1} \bigg).
\label{eq:achievable-rate}
\end{align}
It is revealed from \eqref{eq:achievable-rate} that the optimum choice for the beamforming $\mathbf{v}$ is significantly related to the CSI $\mathbf{h}$. There have been intensive researches on channel training, estimation, and feedback procedures to obtain the forward channels $\mathbf{h}$ at the BS. With the CSI at hand, beamforming solutions can be computed from existing optimization techniques. However, in practical wireless systems, the perfect knowledge of $\mathbf{h}$ is not viable due to errors occurred in the channel acquisition steps, and thus the rate \eqref{eq:achievable-rate} cannot be calculated in optimization processes. It is essential to develop efficient beamforming strategies that are robust to unknown and random CSI errors.

A generic approach to characterize impairments in practical channel acquisition processes is to include a random error $\mathbf{e}_{k}\in\mathbb{C}^{M}$ to the actual channel $\mathbf{h}_{k}$. The corresponding erroneous, or nominal, CSI $\hat{\mathbf{h}}_{k}\in\mathbb{C}^{M}$ is written by \cite{Aquilina:TC15, Li:WCNC07}
\begin{align}
\hat{\mathbf{h}}_{k} = \mathbf{h}_{k} + \mathbf{e}_{k}.
\label{eq:imperfect CSI}
\end{align}
Although the exact realization of the random variable $\mathbf{e}_{k}$ is unavailable, its statistical features, e.g., probability distribution and moments, can be obtained by analytical analysis or measurement processes \cite{Choi-et-al:TWC20}. We denote $\boldsymbol{\epsilon}\in\mathbb{R}^{E}$ of length $E$ as a collection of error statistics, e.g., the covariance of the error vector, that are available at the BS. For instance, in time-division duplex systems, the BS can estimate the CSI using the linear minimum mean square error (MMSE) channel estimation techniques with uplink pilot signal transmission. The estimation error becomes the zero-mean Gaussian vector whose error covariance depends on the uplink signal-to-noise ratio (SNR), pilot designs, and the location of the UEs \cite{Biguesh:TSP06}. These known quantities collectively form the error statistic feature $\boldsymbol{\epsilon}$. Also, in frequency-division duplex protocol, the quality of channel feedback, e.g., the number of feedback bits, acts as the error statistics \cite{Jindal:TIT06}.

In the rest of the section, we formulate a generic beamforming optimization task with imperfect CSI knowledge. To this end, we investigate the information available at the BS. First, the stacked erroneous CSI vector $\hat{\mathbf{h}}\triangleq\{\hat{\mathbf{h}}_{k}:\forall k\in\mathcal{K}\}$ is obtained through a certain channel acquisition process. The error statistics $\boldsymbol{\epsilon}$ can be also known at the BS. Along with the transmit power constraint $P$, the robust beamforming optimization can be defined as a mapping from a three-tuple $(\hat{\mathbf{h}},\boldsymbol{\epsilon},P)$ to the corresponding beamforming $\mathbf{v}$ denoted by
\begin{align}
    \mathbf{v}=\mathcal{V}(\hat{\mathbf{h}},\boldsymbol{\epsilon},P),\label{eq:V}
\end{align}
where $\mathcal{V}:\mathbb{C}^{MK+E+1}\rightarrow\mathbb{C}^{MK}$ indicates a beamforming strategy to be optimized. Using \eqref{eq:V}, the rate of UE $k$ in \eqref{eq:achievable-rate} can be re-expressed as $R_{k}(\mathbf{h}_{k},\mathcal{V}(\hat{\mathbf{h}},\boldsymbol{\epsilon},P))$.

This paper aims at maximizing the sum rate performance by identifying the beamforming optimization policy in \eqref{eq:V}. We consider the average performance expected over the distribution of the actual CSI $\mathbf{h}$ that includes the randomness of the erroneous CSI $\hat{\mathbf{h}}_{k}$ and the error $\mathbf{e}_{k}$. Furthermore, it is required to include the stochastic properties of the error vector $\boldsymbol{\epsilon}$ as well as the power budget $P$ since they affect the computation of the beamforming solution. In particular, if the MMSE channel estimation is adopted, the resulting error statistic feature $\boldsymbol{\epsilon}$, which collects the error variance, changes as the long-term channel statistics vary, e.g., the location of the UEs. In addition, the transmit power budget $P$ can be regarded as a stochastic number \cite{Kim-et-al:WCL20} whose choice depends on the network deployment. Let $\mathcal{E}$ and $\mathcal{P}$ denote the set of all possible values of the error statistic $\boldsymbol{\epsilon}$ and the power budget $P$, respectively. The corresponding formulation is then written~by
\vspace{-4mm}
\begin{subequations}\label{eq:problem-original}
\begin{align}
\underset{\mathcal{V}(\cdot)}{\mathrm{max}}\,\, & \,\mathbb{E}_{\mathbf{h},\boldsymbol{\epsilon},P}\big[\mathcal{R}(\mathbf{h},\mathcal{V}(\hat{\mathbf{h}},\boldsymbol{\epsilon},P))\big]\label{eq:problem-SR-maximization}\\
\mathrm{s.t.}\,\,\,\,\, & \sum\nolimits_{k\in\mathcal{K}}\|\mathbf{J}_{k}\mathcal{V}(\hat{\mathbf{h}},\boldsymbol{\epsilon},P)\|^{2}= P,\ \forall \boldsymbol{\epsilon}\in\mathcal{E},\forall P\in\mathcal{P},\label{eq:constraint}
\end{align}
\end{subequations}
where $\mathbb{E}_{X}[\cdot]$ indicates the expectation evaluated for a random variable $X$, $\mathcal{R}(\mathbf{h},\mathcal{V}(\hat{\mathbf{h}},\boldsymbol{\epsilon},P))\triangleq\sum_{k\in\mathcal{K}}R_{k}(\mathbf{h}_{k}, \mathcal{V}(\hat{\mathbf{h}},\boldsymbol{\epsilon},P))$ denote the sum rate, and $\mathbf{J}_{k}\in\mathbb{R}^{MK\times MK}$ stands for an all zero matrix whose $(M(k-1)+1)$-th to $(Mk)$-th diagonal elements being replaced with ones. The objective function \eqref{eq:problem-SR-maximization} evaluates the conditional expectation of the sum rate which can be achieved over the actual channel distribution for a given erroneous observation $\hat{\mathbf{h}}$ at the BS \cite{Choi-et-al:TWC20}. Therefore, (4) can identify efficient robust operator $\mathcal{V}(\cdot)$ that can compute beamforming vectors well-matched with the unobservable actual channel $\mathbf{h}$ based only on its erroneous observation $\hat{\mathbf{h}}$~\cite{Shen-et-al:TVT14}.

The major obstacle for handling \eqref{eq:problem-original} arises from the mismatch between the actual wireless propagation environment and its imperfect observations required for the robust beamforming calculator $\mathcal{V}(\cdot)$. The mapping $\mathcal{V}(\hat{\mathbf{h}},\boldsymbol{\epsilon},P)$ should fit to the unavailable perfect CSI $\mathbf{h}$ only with the noisy estimation $\hat{\mathbf{h}}$ and the partial information $\boldsymbol{\epsilon}$. For this reason, the optimization process for $\mathcal{V}(\cdot)$ generally relies on the unavailable perfect CSI, which is not viable since the rate \eqref{eq:achievable-rate} cannot be computed without $\mathbf{h}$. Furthermore, the objective function has no analytical expressions even with a simple Gaussian CSI error case \cite{Choi-et-al:TWC20}. To overcome this difficulty, \cite{Choi-et-al:TWC20} derived an approximate sum rate performance by treating the desired signal conveyed through the CSI error as noise. Such an assumption leads to the mismatch between the approximated objective function and the actual achievable rate.
It is necessary to develop a new optimization strategy that directly addresses the intractable objective \eqref{eq:problem-SR-maximization}. To this end, we exploit the data-driven optimization capability of the DL techniques that designs the robust operator $\mathcal{V}(\cdot)$ with numerous training samples.

\vspace{-2mm}
\section{{Deep Learning for Robust Beamforming} \label{sec:Deeplearning-approach}}

\begin{figure}
\centering\includegraphics[width=6.2cm,height=2.4cm]{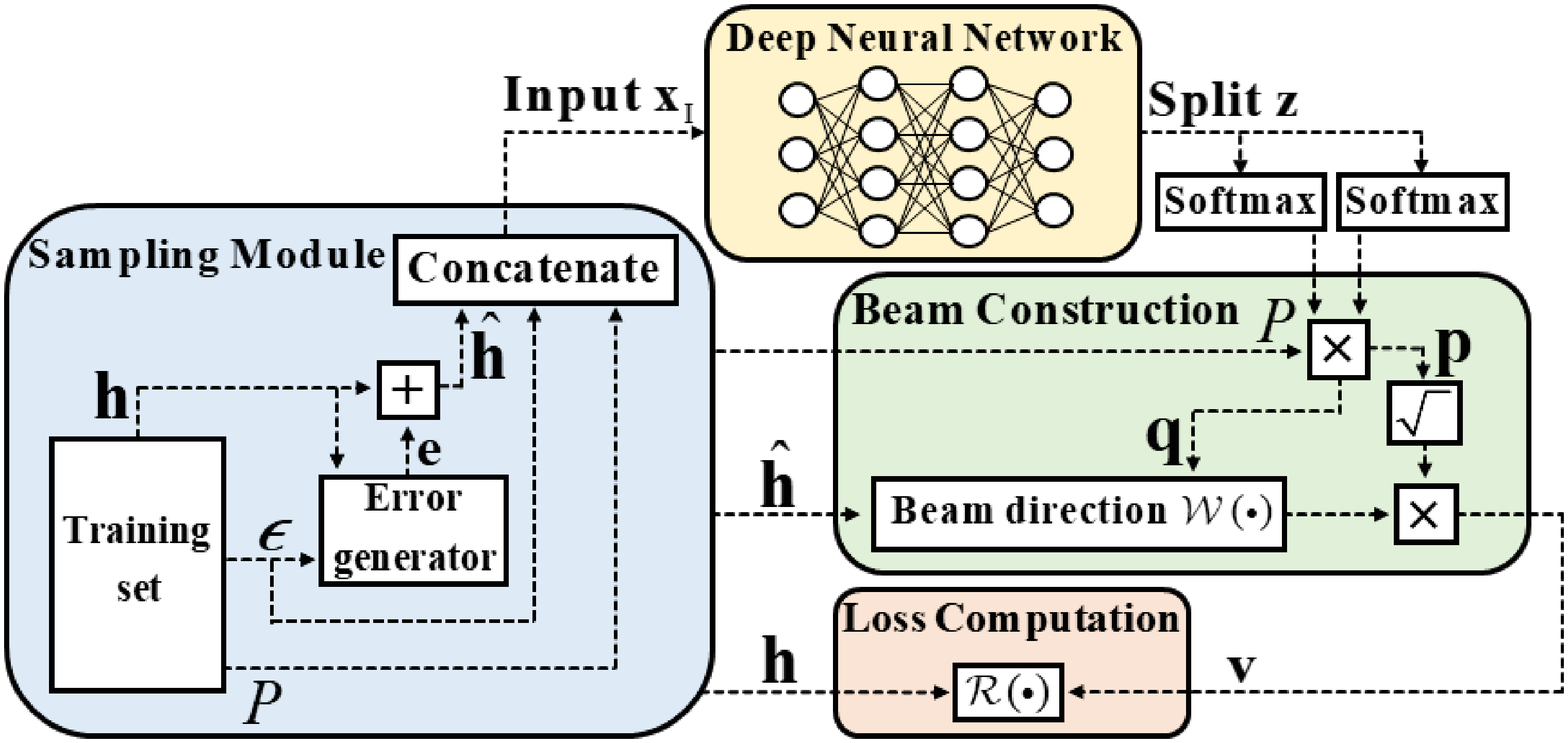} 
\caption{{\label{fig:DNN_structure}Proposed DL framework for robust beamforming.}}
\vspace{-6mm}
\end{figure}

The unstructured mapping $\mathcal{V}(\cdot)$ makes problem \eqref{eq:problem-original} intractable since it requires a search process over infinite-dimensional function spaces. To resolve this issue, we propose a DL-based robust beamforming framework in Fig. \ref{fig:DNN_structure} which replaces the intractable operator $\mathcal{V}(\cdot)$ in \eqref{eq:V} with a carefully-designed DNN. The optimality of such a DNN approximation is verified from the universal approximation theorem \cite{Sun-et-al:TSP18, Hornik-et-al:NN89}. In the following, we first design a DNN architecture that only exploits the imperfect CSI. It is then followed by a robust training procedure.

\vspace{-2mm}
\subsection{{Robust DNN Design}\label{subsec:DNN-structure}}
Let $\mathcal{V}_{\theta}(\cdot)$ be a DNN with a trainable parameter set $\theta$ that approximates the beamforming computation rule $\mathcal{V}(\cdot)$. Unlike existing non-robust DL approaches \cite{Xia-et-al:TC19,Kim-et-al:WCL20} where DNNs are allowed to accept the perfect CSI $\mathbf{h}$, the proposed DNN only observes the estimated channels for the robust design. To this end, an input feature $\mathbf{x}_{\text{I}}$ is designed as a concatenation of the imperfect CSI $\hat{\mathbf{h}}$ and other observable information $\boldsymbol{\epsilon}$ and $P$ and is denoted by $\mathbf{x}_{\text{I}}=[\hat{\mathbf{h}}^{T},\boldsymbol{\epsilon}^{T},P]^{T}\in\mathbb{C}^{MK+E+1}$. The DNN output $\mathbf{x}_{\text{O}}$ is utilized as the beam weights, i.e., $\mathbf{x}_{\text{O}}=\mathbf{v}\in\mathbb{C}^{MK}$. The computation of the DNN  $\mathcal{V}_{\theta}(\cdot)$ is expressed~as
\vspace{-5mm}
\begin{align}
    \mathbf{v}&=\mathbf{x}_{\text{O}}=\mathcal{V}_{\theta}(\mathbf{x}_{\text{I}})\nonumber\\
    &=f_{L}(\mathbf{W}_{L}\times\cdots\times f_{1}(\mathbf{W}_{1}\mathbf{x}_{\text{I}}+\mathbf{b}_{1})+\cdots+\mathbf{b}_{L}),\label{eq:forward-computation}
\vspace{-2mm}
\end{align}
where $f_{l}(\cdot)$ is an activation function of layer $l$ {($l=1,\cdots,L$)}, $\mathbf{W}_{l}\in\mathbb{R}^{N_{l-1}\times N_{l}}$ and $\mathbf{b}_{l}\in\mathbb{R}^{N_{l}}$ respectively denote an weight and a bias of layer $l$, which collectively form a trainable DNN parameter set $\theta\triangleq\{\mathbf{W}_{l},\mathbf{b}_{l}:\forall l\}$, and $N_{l}$ is the output dimension of layer $l$. The activations of the hidden layers, i.e., layers $l=1,\cdots,L-1$, are set to the rectified linear unit (ReLU). The output activation $f_{L}(\cdot)$ for layer $L$ should be carefully chosen to guarantee the power constraint \eqref{eq:constraint}.

We adopt the feature learning structure \cite{Xia-et-al:TC19,Kim-et-al:WCL20} where the output activation function is developed based on the uplink-downlink duality \cite{Bjornson-et-al:SPM14}. The optimal beamforming structure is characterized by a virtual uplink system specified by uplink power variable $q_{k}\geq0$ for UE $k$ under the sum power budget $\sum_{k\in\mathcal{K}}q_{k}=P$ identical to the actual downlink system. Assuming that the perfect CSI is available, the optimal beamforming is written as $\mathbf{v}_{k}=\sqrt{p_{k}}\mathcal{W}_{k}(\mathbf{h},\mathbf{q})$, where $\mathbf{q}\triangleq[q_{1},\cdots,q_{K}]^{T}\in\mathbb{R}^{K}$, $p_{k}\geq0$ stands for the downlink transmit power of UE $k$ under the sum power constraint $\sum_{k\in\mathcal{K}}p_{k}=P$. Here, the function $\mathcal{W}_{k}(\cdot)$ determines the direction of the beam $\mathbf{v}_{k}$ and is defined as
\vspace{-2mm}
\begin{align}
    \mathcal{W}_{k}(\mathbf{h},\mathbf{q})\!\triangleq\!\frac {(\mathbf{I}_{M}\!+\!\sum_{j\in\mathcal{K}}\!q_{j}\mathbf{h}_{j}\mathbf{h}_{j}^{H})^{-1}\mathbf{h}_{k}} {\|(\mathbf{I}_{M}\!+\!\sum_{j\in\mathcal{K}}\!q_{j}\mathbf{h}_{j}\mathbf{h}_{j}^{H})^{-1}\mathbf{h}_{k}\|},\label{eq:Optimal-BF-structure}
\end{align}
with $\mathbf{I}_{M}$ being the identity matrix of size $M$-by-$M$.

The output activation of the proposed robust DNN is designed based on the optimal beamforming structure \eqref{eq:Optimal-BF-structure}.\footnote{Although \eqref{eq:Optimal-BF-structure} might lose the optimality in the erroneous CSI case, it has been shown to be more effective than other types of activations \cite{Kim-et-al:WCL20}.} 
As illustrated in Fig. \ref{fig:DNN_structure}, it consists of two subsequential modules: softmax module and beam construction module. The output of the DNN $\mathbf{z}\triangleq[\mathbf{z}_{p}^{T},\mathbf{z}_{q}^{T}]^{T}$ is set to be a $2K$-dimensional real-valued vector. We first split $\mathbf{z}$ into two $K$-dimensional vectors $\mathbf{z}_{p}$ and $\mathbf{z}_{q}$, each of which is fed to distinct softmax functions. These are then multiplied by the sum power budget $P$ sampled from the training set. The resulting outputs are utilized as the downlink power $\mathbf{p}\triangleq[p_{1},\cdots,p_{K}]^{T}$ and the dual uplink power $\mathbf{q}\triangleq[q_{1},\cdots,q_{K}]^{T}$, respectively, where the sum power constraints $\sum_{k\in\mathcal{K}}p_{k}=\sum_{k\in\mathcal{K}}q_{k}=P$ are always satisfied thanks to the softmax functions. The dual uplink power variables are passed to the operator $\mathcal{W}_{k}(\cdot)$ in \eqref{eq:Optimal-BF-structure} of the beam construction module. Since the perfect CSI is not available, we retrieve the beamforming only with the erroneous channels $\hat{\mathbf{h}}$ as $\mathbf{v}_{k}=\sqrt{p_{k}}\mathcal{W}_{k}(\hat{\mathbf{h}},\mathbf{q})$. Stacking each output $\mathbf{v}_{k}$ for $k\in\mathcal{K}$ forms the final output of the DNN in \eqref{eq:forward-computation}.

\setlength{\textfloatsep}{5pt}{
\begin{algorithm}[t]
\caption{Proposed robust training algorithm}\label{Algorithm}
\begin{algorithmic}
\STATE Initialize $\theta^{[0]}$, and set $t=0$ and $\mathcal{R}_{\text{best}}=\infty$.
\REPEAT
\STATE Sample a mini-batch set $\mathcal{H}$ and generate $\hat{\mathbf{h}}$ from \eqref{eq:imperfect CSI}.
\STATE Update the DNN parameter $\theta^{[t]}$ from \eqref{eq:mini-batch-SGD}.
\STATE Evaluate the validation sum rate $\mathcal{R}_{\text{val}}$.
\IF{$\mathcal{R}_{\text{val}}$ $\geq$ $\mathcal{R}_{\text{best}}$}
\STATE Set $\mathcal{R}_{\text{best}}=\mathcal{R}_{\text{val}}$ and save DNN parameter $\theta^{[t]}$.
\ENDIF
\UNTIL{convergence}
\end{algorithmic}
\end{algorithm}
}

\vspace{-2mm}
\subsection{{Training for Robust Implementation}\label{subsec:Training-Implementation}}
The beam structure in \eqref{eq:Optimal-BF-structure} is developed for the perfect CSI, thereby incurring the mismatch in practical imperfect CSI case. To compensate for this impairment, we present a robust training strategy where the DNN $\mathcal{V}_{\theta}(\cdot)$, whose forward pass computations are carried out only by the erroneous CSI $\hat{\mathbf{h}}$, is trained with numerous realizations of the actual channels $\mathbf{h}$. By replacing the unknown operator $\mathcal{V}(\hat{\mathbf{h}},\boldsymbol{\epsilon},P)$ with the DNN $\mathcal{V}_{\theta}(\mathbf{x}_{\text{I}})$, problem \eqref{eq:problem-original} is recast to an identification task of the DNN parameter $\theta$ written as
\begin{align}
\underset{\theta}{\mathrm{max}}\,\, & \,\mathbb{E}_{\mathbf{h},\boldsymbol{\epsilon},P}\big[\mathcal{R}(\mathbf{h},\mathcal{V}_{\theta}(\mathbf{x}_{\text{I}}))\big],\label{eq:train}
\end{align}
where $\mathcal{R}(\mathbf{h},\mathcal{V}_{\theta}(\mathbf{x}_{\text{I}}))$ indicates the sum rate achieved by the beamforming $\mathcal{V}_{\theta}(\mathbf{x}_{\text{I}})$ in \eqref{eq:forward-computation} computed by the DNN.
The power constraint \eqref{eq:constraint} can be ignored in \eqref{eq:train} since the output activation always meets the feasibility of the problem \eqref{eq:problem-original}. The training formulation \eqref{eq:train} can be addressed via the mini-batch stochastic gradient descent (SGD) methods which iteratively updates the DNN parameter $\theta$ using gradients of the objective function evaluated over a mini-batch set. Let $\mathcal{H}$ be the mini-batch set containing $|\mathcal{H}|$ independent realizations of three-tuple $(\mathbf{h},\boldsymbol{\epsilon},P)$. The SGD update at the $t$-th epoch is expressed as
\begin{align}
\theta^{[t]}&=\theta^{[t-1]}+\eta\frac{1}{|\mathcal{H}|}\sum\nolimits_{(\mathbf{h},\boldsymbol{\epsilon},P)\in\mathcal{H}}\!\!\nabla_{\theta}\mathcal{R}(\mathbf{h},\mathcal{V}_{\theta^{[t-1]}}(\mathbf{x}_{\text{I}})), \label{eq:mini-batch-SGD}
\end{align}
where $\theta^{[t]}$ is the DNN parameter calculated at the $t$-th training epoch, $\eta>0$ stands for the learning rate, and $\nabla_{\theta}$ represents the gradient operation with respect to $\theta$. To implement \eqref{eq:mini-batch-SGD}, both the actual and estimated channels are required. This cannot be straightforwardly addressed by the existing DL works \cite{Xia-et-al:TC19, Kim-et-al:WCL20} since they focused on the ideal perfect CSI case. It is thus essential to develop a new training policy for the robust beamforming optimization.

We summarize the proposed robust training process in Algorithm \ref{Algorithm} which is carried out in advance before the real-time communication services. We first prepare the training dataset by collecting numerous realizations of the perfect CSI $\mathbf{h}$, error statistics $\boldsymbol{\epsilon}\in\mathcal{E}$, and power budget $P\in\mathcal{P}$. This can be achieved via experimental measurement steps or the known distribution of $\mathbf{h}$ with accurate channel models. At each epoch, the mini-batch set $\mathcal{H}$ is randomly sampled from the training set. To this end, as shown in Fig. \ref{fig:DNN_structure}, the sampling module generates the random error vector $\mathbf{e}\triangleq\{\mathbf{e}_{k}:\forall k\in\mathcal{K}\}$ by using the error statistics $\boldsymbol{\epsilon}$ sampled from the training set. The error vector is then added to the actual channel $\mathbf{h}$ to obtain the erroneous CSI. The resulting $\hat{\mathbf{h}}$ is applied to the DNN $\mathcal{V}_{\theta}(\cdot)$ together with the error statistics $\boldsymbol{\epsilon}$ and the power constraint $P$. The forward pass calculations in \eqref{eq:forward-computation} produce the beamforming output $\mathbf{v}$. Then, we can evaluate the sum rate $\mathcal{R}(\mathbf{h},\mathcal{V}_{\theta}(\mathbf{x}_{\text{I}}))$ and its gradient through the backpropagation algorithm. At each training epoch, the generalization capability of the DNN is examined over a validation dataset by computing the validation sum rate $\mathcal{R}_{\text{val}}$. The DNN parameter is saved whenever the current DNN improves $\mathcal{R}_{\text{val}}$. As a result, we can obtain the best DNN parameter with the maximum validation performance.

The proposed training algorithm can be performed in an unsupervised manner without requiring any knowledge regarding the ground truth, i.e., the optimal solution to the original problem \eqref{eq:problem-original}. The proposed DNN observes numerous samples of the perfect CSI $\mathbf{h}$ as well as the corresponding erroneous measurement $\hat{\mathbf{h}}$. Such a data-driven optimization successfully fits the DNN parameter to the distribution of the actual channels even if they are not available in the real-time inference. Furthermore, the proposed robust DNN adopts the error statistic as the side information so that the resulting beamforming becomes resilient for the random changes occurred in the channel acquisition process. This is not viable in the traditional robust optimization algorithms \cite{Wang-et-al:WCL12, Choi-et-al:TWC20} developed for the fixed error statistics.

Once $\theta$ is determined, the beamforming vector for a new estimated channel $\hat{\mathbf{h}}$ can be obtained from \eqref{eq:forward-computation}. Hence, the online computations of the trained DNN can be implemented without the knowledge of the perfect channels by means of the trained parameter set $\theta$ stored in the BS. The computational complexity of the trained DNN relies on its structure, e.g., the number of layers and the beam recovery process \eqref{eq:Optimal-BF-structure}. Since the structure of the DNN is fixed as constants, the matrix inversion in \eqref{eq:Optimal-BF-structure} dominates the overall complexity. Consequently, the computational complexity is given as $\mathcal{O}(KM^{2}+M^{3})$.

\vspace{-2mm}
\section{Numerical Results}\label{sec:numerical_results}

\begin{figure}
\centering\includegraphics[height=.4\linewidth]{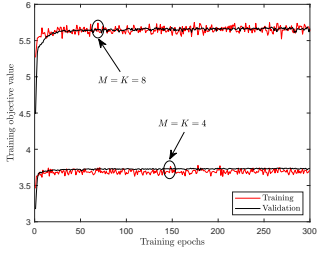}
\caption{\label{fig:graph1}The training and validation convergence versus epochs for $M=K\in\{4,8\}$.}
\end{figure}

We demonstrate the effectiveness of the proposed robust DL approach via numerical simulations. The UEs are uniformly distributed within a circle area of radius $100\text{m}$ and the BS is located at the center of circle. The Rayleigh fading is considered as $\mathbf{h}_{k}\sim\mathcal{CN}(\mathbf{0}, \sqrt{\rho_{k}}\mathbf{I}_{M})$ where $\rho_{k}\triangleq1/(1+(d_{k}/d_{\text{ref}})^{\alpha})$ is the long-term pathloss, $d_{k}$ represents distance between the BS and UE $k$, $d_{\text{ref}}=30\text{m}$ indicates the reference distance, and $\alpha=3$ is the path-loss exponent. The power budget $P$ is uniformly distributed over $P\in\{0\ \text{dB},5\ \text{dB},\cdots,30\ \text{dB}\}$. Assuming the unit noise variance, the transmit SNR is defined as $P$. Elements of the error vector $\mathbf{e}_{k}\sim\mathcal{CN}(\mathbf{0},\epsilon_{k}\mathbf{I})$ follows the Gaussian distribution with zero mean and variance $\epsilon_{k}$. Unless stated otherwise, the error statistic is defined as $\boldsymbol{\epsilon}\triangleq\{\epsilon_{k}:\forall k\in\mathcal{K}\}$ with the length $E=K$. We set $\epsilon_{k} = \tau||\mathbf{h}_{k}||_{2}^{2}$ where $\tau\in(0,1]$ stands for the error ratio factor indicating the fraction of the CSI error in terms of the channel gain $\|\mathbf{h}_{k}\|^2_{2}$. The error ratio is uniformly generated as $\tau \in \{0.005, 0.01, 0.05, 0.1, 0.3, 1\}$. Five fully-connected hidden layers each with $20MK$ output dimension are examined. The Adam optimizer is applied with learning rate $\eta=0.001$ and $10^{4}$ mini-batch samples. The batch normalization is adopted at hidden layers. The performance of the trained DNN is evaluated with $10^{3}$ test samples.

The convergence behavior of the proposed robust training algorithm is presented in Fig. \ref{fig:graph1} which evaluates the objective function \eqref{eq:train} over the training and validation samples with respect to the training epochs. The objective value gradually increases as the DNN gets trained. This validates the effectiveness of the proposed unsupervised training policy.

\begin{figure}
\begin{minipage}[t]{0.45\columnwidth}%
\centering\includegraphics[width=4.3cm,height=4cm]{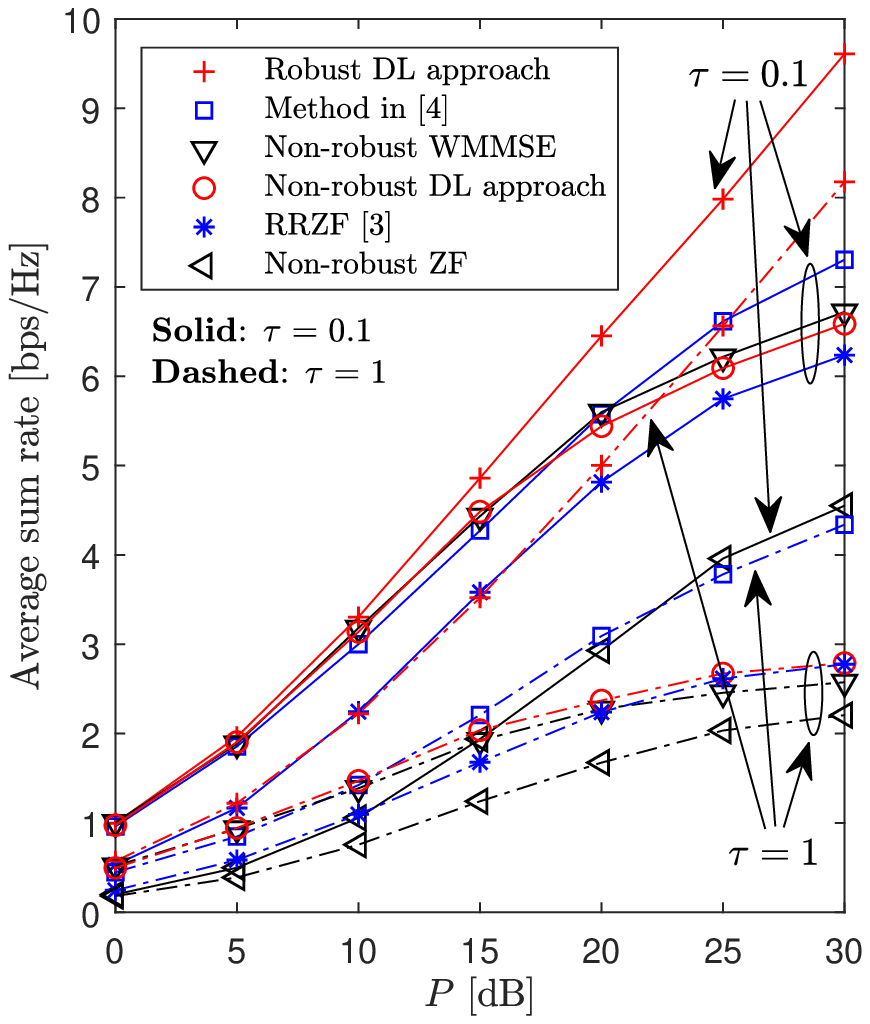}

\centering~~~{\scriptsize{}(a) $M=K=4$}{\footnotesize \par}%
\end{minipage}~~~~%
$\quad\!\!\!\!\!\!$
\begin{minipage}[t]{0.45\columnwidth}%
\centering\includegraphics[width=4.3cm,height=4cm]{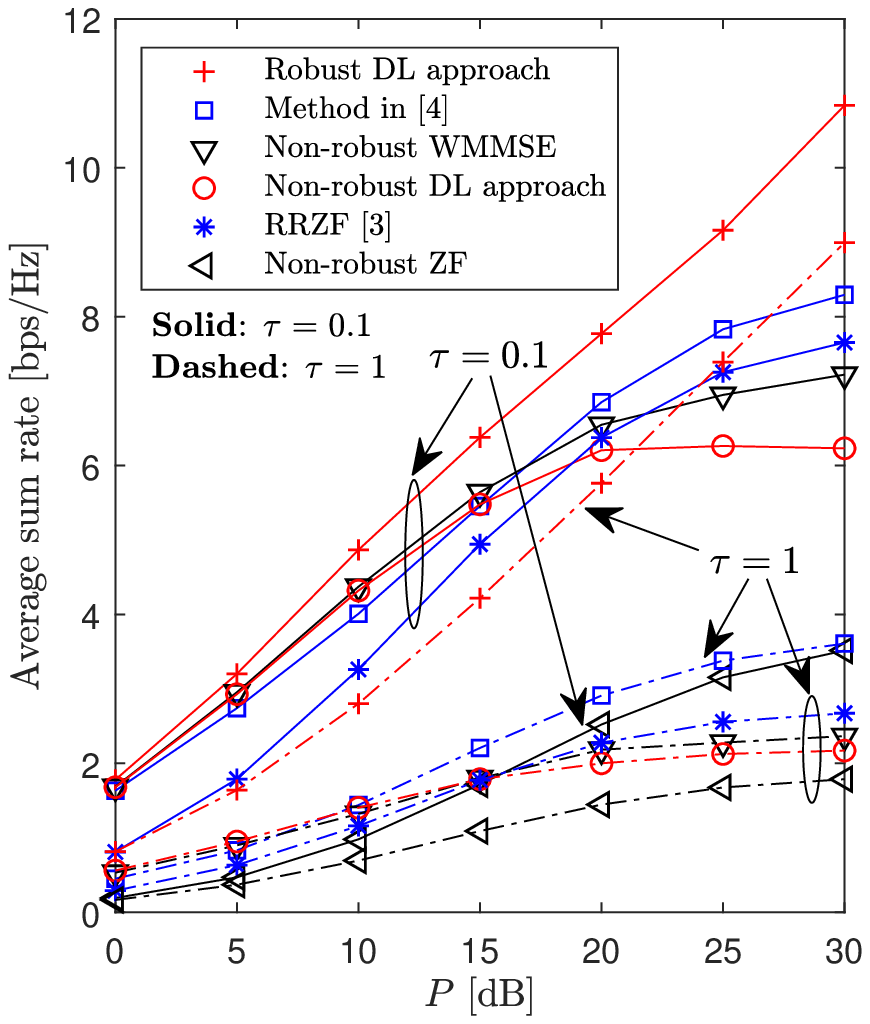}

\centering~~~{\scriptsize{}(b) $M=K=8$}{\footnotesize \par}%
\end{minipage}

\caption{\label{fig:graph3}{Average sum rate versus $P$ for $M=K\in\{4,8\}$.}}
\vspace{-6mm}
\end{figure}

\begin{figure}
\begin{minipage}[t]{0.45\columnwidth}%
\centering\includegraphics[width=4.3cm,height=4cm]{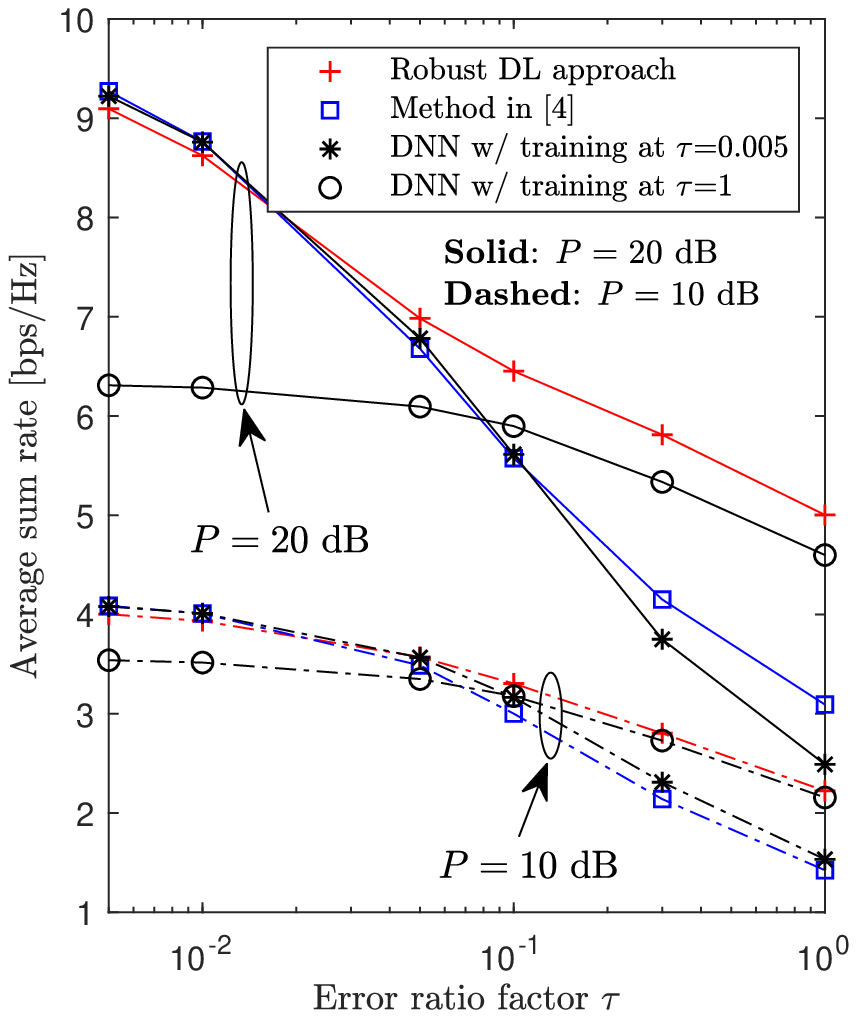}

\centering~~~{\scriptsize{}(a) $M=K=4$}{\footnotesize \par}%
\end{minipage}~~~~%
$\quad\!\!\!\!\!\!$
\begin{minipage}[t]{0.45\columnwidth}%
\centering\includegraphics[width=4.3cm,height=4cm]{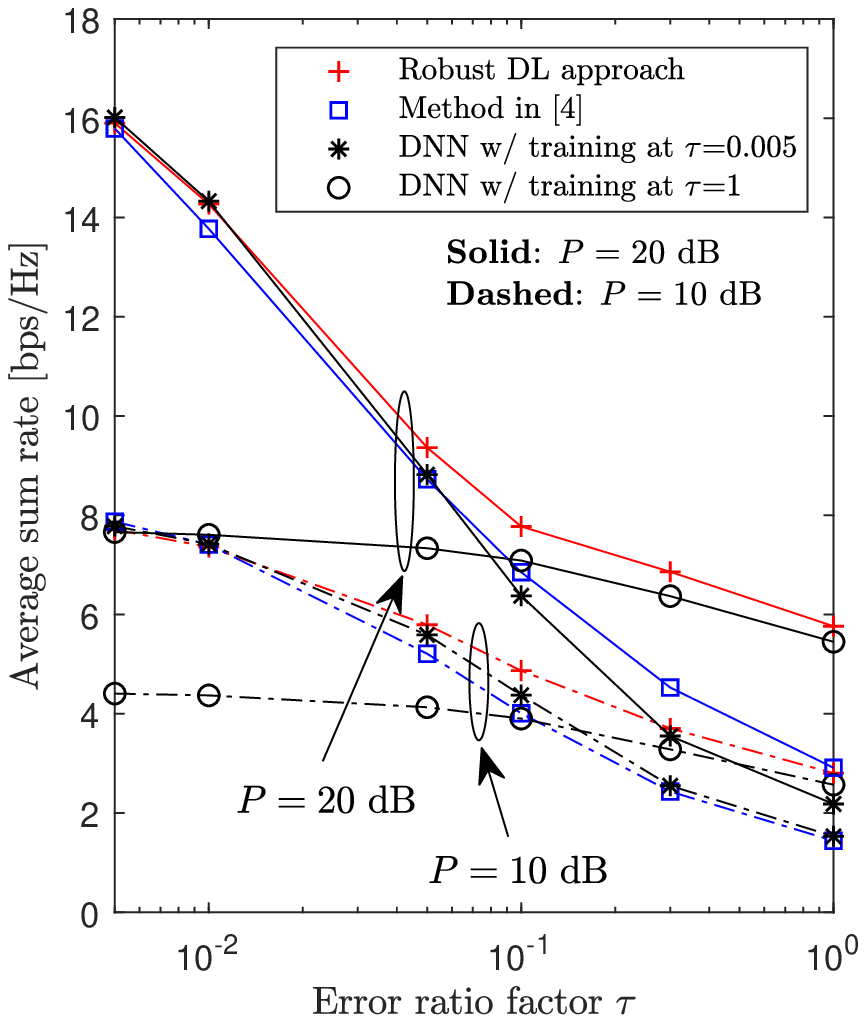}

\centering~~~{\scriptsize{}(b) $M=K=8$}{\footnotesize \par}%
\end{minipage}

\caption{\label{fig:graph4}{Impact of the robustness for $M=K\in\{4,8\}$.}}
\end{figure}

Fig. \ref{fig:graph3} shows the average sum rate versus the SNR for $M=K\in\{4,8\}$ with $\tau = 0.1$ and $1$. As a benchmark, the conventional non-robust DL \cite{Xia-et-al:TC19,Kim-et-al:WCL20} is considered which trains a DNN only with the perfect CSI. We also plot the performance of traditional non-robust transmission methods, i.e., the WMMSE \cite{Christensen-et-al:TWC08} and the ZF, as well as robust schemes such as the RRZF \cite{Wang-et-al:WCL12} and the iterative beamforming optimization \cite{Choi-et-al:TWC20} developed for the imperfect CSI case. All the baseline schemes need to optimize the beamforming vectors for each realization of $\boldsymbol{\epsilon}$ and $P$. On the other hand, the proposed robust DL approach directly identifies the computation rule for any given $\boldsymbol{\epsilon}$ and $P$. Nevertheless, regardless of the network size, it outperforms the baseline methods for all simulated setups, especially at high $\tau$ and $P$. We see a substantial performance gain over the baseline methods. The proposed robust DL scheme directly tackles intractable objective function \eqref{eq:problem-SR-maximization} via the data-driven optimization strategy \eqref{eq:mini-batch-SGD}. In contrast, the method in \cite{Choi-et-al:TWC20} maximizes an approximated sum rate function since the traditional algorithm requires closed-form expressions for the objective function. Such a model mismatch leads to the performance degradation of the existing approaches. It is observed that the performance of the baseline schemes saturate at the high SNR regime due to the residual interference power induced by the channel estimation error. Such an issue can be successfully managed by the proposed robust DL scheme whose sum rate monotonically grows with $P$. This verifies the robustness of the proposed method to the erroneous channel acquisition processes with arbitrary error variance and the SNR.

Fig. \ref{fig:graph4} depicts the average sum rate as a function of the error ratio factor $\tau$ with $M=K\in\{4,8\}$. To see the impact of the error variance, we exhibit the performance of DNNs trained at specific error ratios $\tau=0.005$ and $1$ but tested over arbitrary $\tau$. Regardless of $P$, the proposed robust DNN trained with various $\tau$ performs well for all simulated $\tau$ and outperforms the DNNs each trained at a certain error ratio. This indicates that the proposed training strategy, which exploits the error variance $\boldsymbol{\epsilon}$ as side information, is crucial for improving the sum rate performance and the generalization ability. Also, we observe that the method in \cite{Choi-et-al:TWC20} incurs severe performance loss in the high $\tau$ regime, i.e., when the channel estimation becomes inaccurate. Since the actual channels are not available, the method in \cite{Choi-et-al:TWC20} maximizes an approximated metric of the sum rate which can be evaluated with the estimated CSI $\hat{\mathbf{h}}$ and the error variance $\boldsymbol{\epsilon}$. However, the approximation becomes inaccurate as the error variance gets larger. Such a model mismatch can be resolved by the proposed robust DNN since it can get the information regarding the exact average sum rate performance in the training step. These results validate the generalization ability and the robustness of the proposed DL approach for arbitrary error variance.

\begin{figure}
\centering\includegraphics[height=.4\linewidth]{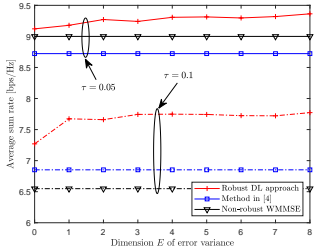}
\caption{{\label{fig:graph2}The average sum rate versus dimension $E$ of error variance for $M=K=8$.}}
\vspace{-2mm}
\end{figure}

\begin{table}[]
\centering\caption{\label{tab:table1}Average CPU Running Time [sec].}
\renewcommand{\tabcolsep}{1.7mm}
\begin{tabular}{cccccccc}
\hline
\multirow{2}{*}{} & \multirow{2}{*}{\begin{tabular}[c]{@{}c@{}}Robust DL\\ approach\end{tabular}} & \multicolumn{2}{c}{Non-robust WMMSE}      & \multicolumn{2}{c}{Method in [4]}                \\  \cline{3-6}
                  &
                   & $P=0$    &  $P=30$       & $P=0$    &     $P=30$        \\ \hline
$M=K=4$                 & 3.52e-4                      & 1.77e-2     &     5.24e-1                & 1.88e-2   &      3.96e-1        \\ \hline
$M=K=8$                 & 2.68e-3                       & 3.28e-2    &      1.37e-0           & 3.84e-1    &     6.16e-1   \\ \hline
\end{tabular}
\end{table}

To see the impact of the error statistic input $\boldsymbol{\epsilon}$, Fig. \ref{fig:graph2} evaluates the average sum rate by varying the dimension $E$ of $\boldsymbol{\epsilon}$ for $M=K=8$ with $\tau\in\{0.05,0.1\}$ and $P=20$ dB. The BS is assumed to know the error variance of randomly selected $E\leq K$ UEs. Thus, the DNN can only exploit the partial error variances as the side input. The performance of the method in \cite{Choi-et-al:TWC20}, which requires the full knowledge of the error variances of all $K$ UEs, is also plotted. The performance of the proposed scheme monotonically increases as the dimension $E$ grows, implying that the side information $\boldsymbol{\epsilon}$ is beneficial for the proposed robust DNN. The proposed robust DL is superior to the conventional scheme \cite{Choi-et-al:TWC20} regardless of $E$. This demonstrates that the proposed robust training policy is effective even when we can only get access to the partial information of the error statistics.

Finally, we present the time complexity of various schemes in Table I for $M=K\in\{4,8\}$ with $\tau=1$. Regardless of the network size, the proposed robust DL framework exhibits much lower computation time than the conventional methods. This is because, the computation of the DNN depends on simple forward pass computation \eqref{eq:forward-computation}, whereas the conventional schemes operate iteratively for each given channel realizations. Also, we can observe that the CPU execution time of the conventional methods increases with the SNR $P$ as they require a more number of iterations for the convergence. On the contrary, the proposed DL approach guarantees the identical time complexity for all $P$ since the DNN structure remains the same. Thus, we can conclude that the robust DL framework is powerful both in terms of the sum rate performance and the time complexity.

\vspace{-1mm}
\section{Conclusion\label{sec:Conclusion}}
This work has developed the DL framework for determining robust beamforming solutions for the downlink MISO systems. The DNN is designed to process with the erroneous CSI as well as its error statistics. To narrow the mismatch between the input erroneous CSI and the actual channel, we propose a robust training strategy which optimizes the DNN by using both the perfect and imperfect CSIs. Numerical results verify the superiority of the proposed robust DL approach over conventional beamforming schemes.

\vspace{-1mm}
\renewcommand{\baselinestretch}{0.90}

\end{document}